\documentstyle[twoside,fleqn,epsf,espcrc2]{article}

\newcommand{\AmS}{{\protect\the\textfont2
  A\kern-.1667em\lower.5ex\hbox{M}\kern-.125emS}}

\title{ Comparison of Multi-quark Matrix Inversion Algoritmes}

\author{He-Ping Ying,\address{Zhejiang Institute of Modern Physics, 
        Zhejiang University, Hangzhou 310027, P.R China}%
	$^,$\address{Department of Physics and Astronomy, University 
	of Kentucky, Lexington KY, 40506-0055, USA}~
        Shao-Jing Dong${\rm ^b}$
        and 
        Keh-Fei Liu$^{\rm b}$}
       
\begin{document}

\begin{abstract}
We test iterative algorithms, MR, QMR$\gamma_5$ and BiCG$\gamma_5$,
to compare their efficiency in matrix inversion with multi-quarks
{\it (shifted matrices)} within one iteration process. Our results
on the $8^3 \times 12$ and $16^3 \times 24$ show 
that MR admits multi-quark calculation with less memory requirement,
whereas QMR is faster for the single quark calculation. 
\end{abstract}

\maketitle

\section{INTRODUCTION}
To describe the inversion algorithms for multi--quarks, we start by giving 
a general formula for {\it shifted} matrix,
a multiple of the identity plus a constant off-diagonal part,
\begin{eqnarray}
A(\sigma) =  \sigma {\bf 1} + A.~~~~~~~~~~~~~~~~~~~~~~~~~~~~~
\end{eqnarray}
The parameter $\sigma$ stands for a whole trajectory.  
In lattice QCD theory the fermion matrix,
with respect to quark mass\cite{Gupta89} 
\begin{eqnarray}
m=\frac{1}{2a}(\frac{1}{\kappa} - \frac{1}{\kappa_c}),
\end{eqnarray}
has the shifted structure\cite{From95}, and they are related to each other by,
\begin{eqnarray}
M_{heavy} = f(m) + M_{light},~~~~~~~~~~~~~
\end{eqnarray}
where term $f(m)=(\kappa^{-1}_{heavy}- \kappa^{-1}_{light}) >0$, $M_{light}$ 
and $M_{heavy}$ are fermion matrices for {\it light} quark (considered as {\it 
seed} system) and {\it heavy} quark ({\it extrapolated} system) respectively.

By numbering all even sites before the odd ones, we rewrite 
Dirac equation $Mx = \phi$ as
\begin{eqnarray}
\left( \begin{array}{cc}
\sigma{\bf 1}  & -D_{eo} \\
-D_{oe}  & \sigma{\bf 1}  \end{array}\right)
\left(\begin{array}{c}
x_e \\
x_o\end{array}\right)
=\left(\begin{array}{c}
\phi_e \\
\phi_o\end{array}\right),
\end{eqnarray}
where $\sigma =1/\kappa$, then separate it into the so-called even-odd 
preconditioned form:
\begin{eqnarray}
&M_e x_e =  \tilde\phi_e,~~~
x_o = \kappa (\phi_o + D_{oe} x_e);~\\
& M_e = \sigma^2 - D_{eo} D_{oe},~~
\tilde\phi_e = \sigma \phi_e + D_{eo}\phi_o.\nonumber
\end{eqnarray}
For the smeared source the equation  $M_e x_e = \tilde \phi_e$
is further decoupled, by setting $x_e = \sigma y_e + z_e$,
\begin{eqnarray}
M_e y_e = \phi_e,~~~
M_e z_e = D_{eo} \phi_o,~
\end{eqnarray}
such that the right hands of these equations are independent of $\kappa$.

An iterative process to solve the nonsingular system $Ax= b$ starts
from an initial guess $x^0$ and an initial residual  $r^0 = b-Ax^0$.
The nonsymmetric Lanczos process\cite{Borici94,From94} generates an 
orthogonal basis for the Krylov subspace
\begin{eqnarray}
K_m(A,r^0)= \{r^0, Ar^0, A^2r^0, \cdots, A^{m-1}r^0\},
\end{eqnarray}
to obtain an approximate solution $x^m$ in $m$th step 
iteratively with short recurrences and to keep $x^m \in x^0 + K_m(A,r^0)$.
It is essential to notice for inversion with multi-quarks that,
on the trajectory of the shifted matrices, their Krylov spaces
are identical\cite{From95,Forc96}.

Two directions to achieve good efficiency,
besides a good preconditioning\cite{Fisch96}, are considered
currently\cite{From95}: (a) Acceleration of convergence using improved 
iterative procedures (such as QMR and BiCGStab2).
(b) Exploitation of structure of the matrix $M$ in the inverters
(such as  $\gamma_5$--symmetry and {\it shifted} properties).
In this paper we attempt to test Minimal Residual (MR), Quasi-Minimal
Residual (QMR) and Bi-Conjugate Gradient (BiCG), exploiting the
$\gamma_5$--symmetry and using the {\it shifted} feature 
for inversion with multi-quarks\cite{Brow95,Glas96,From96}. 
For definiteness we consider only $\delta$--sources and solve one of the  
expressions in eq. 6 as $y_e=0$ or $z_e=0$.

\begin{figure}[htb]
\setlength\epsfxsize{70mm}
\epsfbox{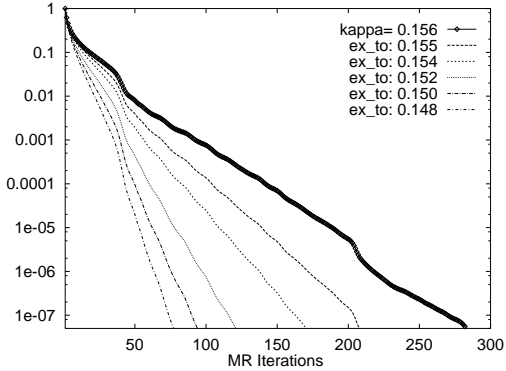}
\caption{The  relative residuals versus iteration using M$^3$R algorithm,
on $8^3\times 12$ lattice.}
\end{figure}
\section{ALGORITHM}
First we mention that, after even-odd precondition, the matrix $M_e$ still has 
$\gamma_5$--symmetry and they are shifted matrices for multi-quarks.
Our numerical computations were done 
for the even-odd preconditioned systems
on lattices $8^3\times 12$, and $16^3\times 24$ and for the quenched gauge 
configurations at $\beta = 6.0$ ($\kappa_c \simeq 0.157$). 
We tested the Multiple-Masses-Minimal Residual (M$^3$R)
method for matrix inversion with multi-quarks 
and then compare the results of convergence rates with those obtained by 
using the QMR and BiCG algorithms.  
\subsection{M$^3$R Algorithm[8] }
To solve $(\sigma {\bf 1} + A)x = b$, the M$^3$R algorithm is given by 
(initial: $x^0=0,~r^0=b-Ax^0,~f_{-1}^\sigma=1$),
\begin{eqnarray*}
{\bf do}~m &=& 0, 1, \cdots, to~convergence ~~~~~~~~~~~~~~~~~~~~~~~\\
 p^m &=& Ar^m\\
\alpha_m &=&  \omega \frac{(p^m)^\dagger r^m}{(p^m)^\dagger p^m}\\
 f^\sigma_m &=&  \frac{f^\sigma_{m-1}}{(1.0 + \sigma \alpha_m)}~\\
x^{m+1} &=&  x^m + \alpha_m r^m\\
x_\sigma ^{m+1} &=&  x_\sigma ^m + \alpha_m f^\sigma_m r^m\\
r^{m+1} &=& r^m - \alpha_m Ar^m\\
{\bf end}~~{\bf do}
\end{eqnarray*}
where $x^m$ and $r^m$ are the $m$th approximate solution and residual
respectively for the seed system. $x^m_\sigma$ is the $m$th approximate 
solution for one of the extrapolated systems and coefficient $f^\sigma_m$
is iterated step by step for each quark mass. It is necessary to take $x^0=0$ 
for seed system to keep all initial residuals $r^0_\sigma$ to be the same for 
different quark masses.  As shown by the algorithm, the matrix-vector 
multiplication 
performs only once in the whole set $\{\sigma\}$ at each iterative step.
For each additional quark mass, the price to pay is one vector
$x^\sigma_m$ to be stored and a little CPU times (about 8\% for scalar
products), with no  additional matrix multiplication performed.
\begin{figure}[htb]
\setlength\epsfxsize{70mm}
\epsfbox{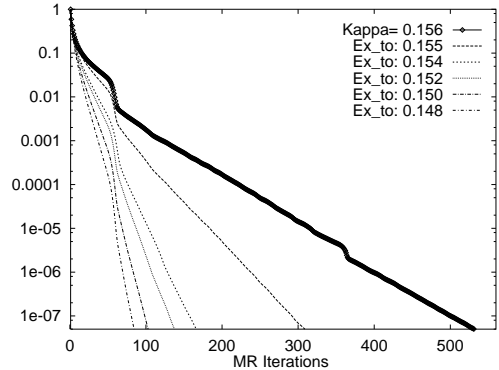}
\caption{The same as Figure 1. but on $16^3\times 24$ lattice.}
\end{figure}
We take the system at $\kappa = 0.156$ as a seed and extrapolate to
heavier quarks at $\kappa = 0.155, 0.154, 0.152, 0.150$ and 0.148, as shown
in Fig. 1 and Fig. 2.  The results give evidence that the gain
factor is about 2 by using the M$^3$R for 5 extrapolated quarks
as compared to calculating the 5 quarks separately for $\delta$ sources.
The overrelaxation parameter $\omega$ is chosen to be $\omega = 1.1$ 
(Fig. 3) for best convergence rate. In these plots the relative residual is 
defined by $\parallel b-Ax^m\parallel/\parallel r^0\parallel$.
The stopping criteria for convergence is $\parallel r^m\parallel\leq 10^{-9}$.
\begin{figure}[htb]
\setlength\epsfxsize{70mm}
\epsfbox{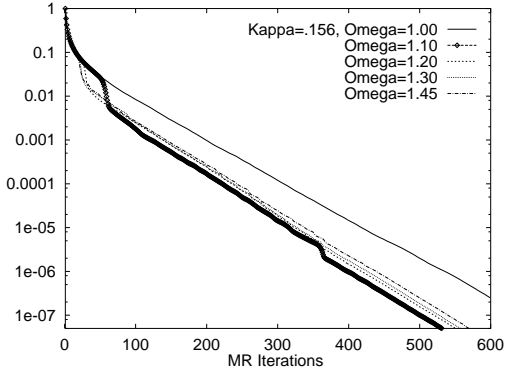}
\caption{The convergence rate in MR for different values of $\omega$ on 
$16^3\times 24$ lattice.}
\end{figure}
\subsection{QMR$\gamma_5$ without look-ahead}
The Quasi-Minimal Residual  exploiting the $\gamma_5$-symmetry 
is described in ref. \cite{From95}. To solve  eq. 6 for shifted matrix
(see eq. 1), it performs:
\begin{eqnarray*}
{\bf do}~m &=& 1, 2, \cdots, to~convergence ~~~~~~~~~~~~~~~~~~~~~~~~~~~~~\\
\{ {\rm I.}&do&{\rm Lanczos ~ step } \}\\
\delta_m &=&  (\gamma_5 \nu^m)^\dagger \nu^m \\
\alpha _m &=&  (\gamma_5 \nu^m)^\dagger A\nu^m/\delta_m + \sigma\\
\beta_m &=& \rho_m \delta_m / \delta_{m-1} \\
r^{m+1} &=&  A \nu^m - (\alpha_m-\sigma)\nu^m - \beta_m \nu^{m-1}\\
\rho_{m+1} &=& \parallel r^m\parallel\\
\nu^{m+1} &=&  r^{m+1}/\rho_{m+1}\\
\{ {\rm II.}&for&{\rm QMR ~~recurrence~~coefficients } \}\\
\{\alpha_m,\beta_m\} &\rightarrow& \{\theta, \varepsilon,
c_{m+1}, s_{m+1}, \chi_{m+1} \} \\
\{ {\rm III.}&do&{\rm QMR~~iterations } \}\\
p^{m}&=&(\nu^m-\varepsilon p^{m-1}-\theta p^{m-2})/\chi_{m+1}\\
\tilde \rho_m &=&  c_{m+1} \rho_m,\\
x^{m+1} &=& x^m+ \tilde \rho_m p^m, \\
\rho_{m+1} &=&  -\bar s_{m+1} \rho_m\\
{\bf end}~{\bf do}
\end{eqnarray*}
\begin{figure}[htb]
\setlength\epsfxsize{70mm}
\epsfbox{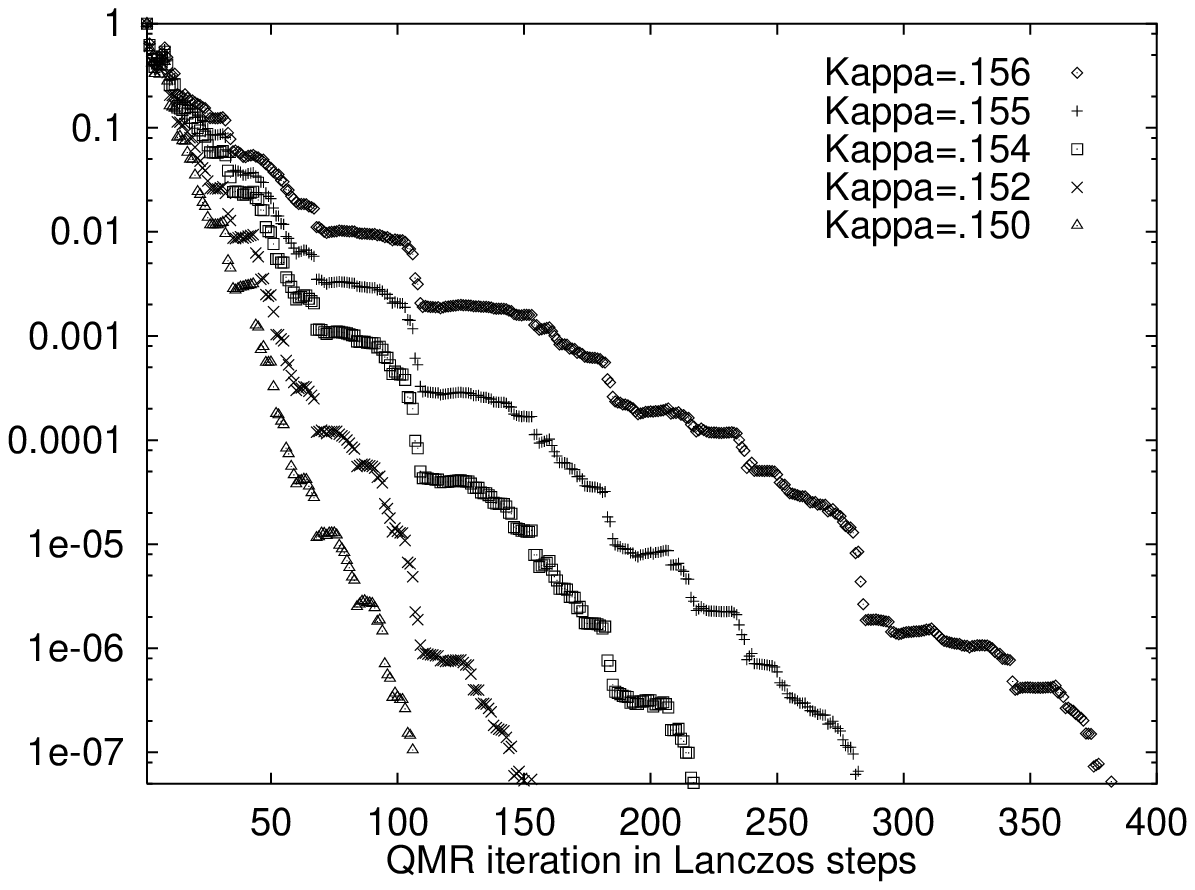}
\caption{The relative residuals versus iteration by using the QMR algorithm, 
on $16^3\times 24$ lattice.}
\end{figure}
\begin{figure}[htb]
\setlength\epsfxsize{70mm}
\epsfbox{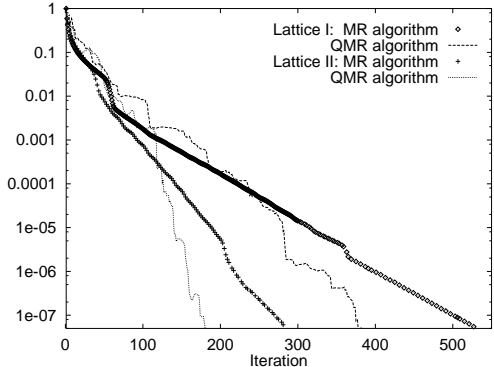}
\caption{The curves of convergency at $\kappa = 0.156$ using MR and 
QMR$\gamma_5$ algorithms, from up to low, on Lattice I: $16^3\times 24$ 
and II: $8^3\times 12$.}
\end{figure}
In steps II and III, there is no matrix multiplication  
in the $m$th approximation.
It is obvious that, to solve the Dirac equation with multi-quarks,
the matrix multiplication is carried out only at $\sigma=0$
point for the whole set of $\sigma$. Due to the $\gamma_5$-symmetry of $M_e$,
$\gamma_5 M_e^\dagger \gamma_5 =M_e$, the computational effort at each Lanczos
step reduces to one matrix multiplication, instead of two (for $M_e$ and 
$M_e^\dagger$ each), and the coefficients are all real.
The price to pay is three vectors $ \{ x^m(\sigma),
p^m(\sigma), p^{m-1}(\sigma) \}$ to be stored for each additional quark.
Fig. 4 gives the results of the relative residuals 
versus the iteration steps at several $\kappa$'s respectively.
The plot of Fig.5  shows that QMR$\gamma_5$ is faster than MR in convergency 
at $\kappa = 0.156$.  But this feature could be reduced by the GMRES(4), 
which can save 30\% iterations compared with MR, but at the expense of 
3 more vectors in memory\cite{From96}.

\subsection{BiCG$\gamma_5$ Algorithm}

BiCG method\cite{Forc96} exploiting the $\gamma_5$-symmetry is
\begin{eqnarray*}
{\bf do}~m &=&0, 1,\cdots,~to~convergence\\
\delta_m &=& (\gamma_5 r^m)^\dagger r^m/(\gamma_5 p^m)^\dagger A(\sigma)p^m\\
x^{m+1} &=& x^m + \delta_m p^m\\
r^{m+1} &=& r^m - \delta_m A(\sigma) p^m\\
\beta_m &=& (\gamma_5 r^{m+1})^\dagger r^{m+1} / (\gamma_5 r^m)^\dagger r^m\\
p^{m+1} &=& r^{m+1} + \beta_m p^m\\
{\bf end}~{\bf do}
\end{eqnarray*}
\begin{figure}[htb]
\setlength\epsfxsize{70mm}
\epsfbox{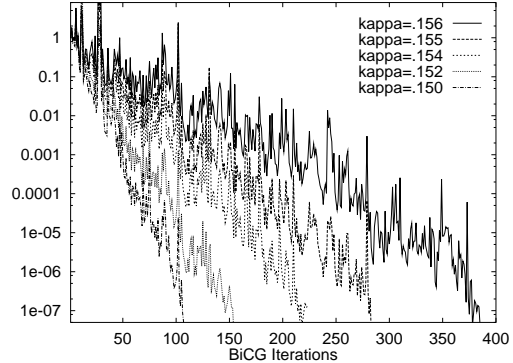}
\caption{The relative residuals versus iteration by using the BiCG algorithm, 
on $16^3\times 24$ lattice.}
\end{figure}
This two-term recurrence method has 
difficulty in memory capacity for multi--quarks: the coefficients in the 
$m${th} step for different values of $\sigma$ can not be obtained
from those for $\sigma=0$ by short recurrences. In addition, this algorithm 
also shows large fluctuations in the relative residual (Fig. 6) 
which can be eliminated 
by the variant BiCGStab\cite{From94} algorithm.

\section{CONCLUSION}
For problems involving the inversion of multi-quark matrices, we find
M$^3$R or GMRES to be a good compromise if the memory is limited. It requires 
only one more vector for each additional quark and the overhead in CPU time is 
minimal ($\sim 8\%$). On the other hand, QMR is faster than M$^3$R.
However, it requires memory of 3 vectors for each additional quark and
a look-ahead algorithm to avoid the breakdown\cite{Freu91}. BiCG$\gamma_5$
does not admit multi-quark implementation with short recurrences and the
relative residual fluctuates in a large range.

\section{ACKNOWLEDGEMENTS}
We would like to thank A. Frommer and \mbox {P. De Forcrand} for helpful 
discussions
and comments, and for providing the preprint. HPY is supported by the Center
for Computational Sciences (CCS) at University of Kentuchy and the NSF of 
China.  He also thanks Z. Bai for helpful discussions.
The calculations were performed on CONVEX C240 at University of Kentuchy.

\end{document}